# R2P2: Reactive Routing and Payment Protocol for Named Data Network using Blockchain

Yuhang Ye*, Brian Lee†, Yuansong Qiao*
Software Research Institute, Athlone Institute of Technology
Athlone, Co. Westmeath, Ireland
*{yye, ysqiao}@research.ait.ie; †blee@ait.ie

*Abstract*—With the continuous emergence of new mobile devices which support new communication paradigms such as D2D and V2V, Internet users can take advantage of these devices to achieve better Internet connectivity and improve service quality. Meanwhile, packet forwarding brings extra costs to devices (e.g. electricity consumption), that hinders the realisation of successful ad-hoc networks. This paper proposes Reactive Routing and Payment Protocol (R2P2) to incentivise mobile devices to contribute idle networking resources and gain monetary returns. The routing and payment protocol is developed for Named-Data Network (NDN) because its content-centric nature can better support the intermittent and ephemeral communication requirements in ad-hoc networks. Blockchain is used as the settlement platform for transactions between devices because of its neutrality, robustness and trust. R2P2 is still an on-going project. The content of this paper focuses on the design of R2P2.

## I. Introduction

The future Internet is evolving into a "hodgepodge". Not only existing infrastructures but also scattered mobile devices will have a chance to participate in networking. According to a report by Juniper Research titled "Consumer Connected Cars: Telematics, In-Vehicle Apps & Connected Car Commerce 2018-2023," [1], by 2023 about 60% of new vehicles (representing over 62 million vehicles) will have vehicle-to-vehicle (V2V) capability. In the future, moving vehicles will potentially formulate ad-hoc networks to support emerging driving applications (e.g. self-driving and traffic planing). The same trend is occurring in handheld smart devices. With device-to-device (D2D) communication [2], the dense and wide coverage of these devices (e.g. smartphones and smartwatches) enables massive connectivities and bandwidth to support user-centric applications (e.g. video sharing).

To enable networking amongst emerging mobile devices thus supporting versatile application services, one of the fundamental challenges is how to incentivise mobile devices to share their idle resources so that more effective connections can be created. It is a true barrier, that routing data via a sequence of mobile devices will result in non-trivial costs such as high battery consumptions and degenerated user experiences. A promising solution has been to tokenise and trade the networking services enabled by idle resource. Because the devices who provide services will gain monetary income from the devices who consumer services, it eliminates free-riders and to certain extent, incentivise resource sharing.

This paper introduces Reactive Routing and Payment Protocol (R2P2) to enable trade of network resources. R2P2 employs Named Data Networking (NDN) as the transmission protocol. Compared with traditional host-centric models (such as IP), NDN can better support device mobility and intermittent connectivity which are the nature in mobile ad-hoc networks. Host-centric fails to fit the mobile environment because it is difficult to assign host addresses to moving nodes. Moreover, the host-centric model is inherently not suitable for emerging applications where the goal is data consuming regardless of who requests and who serves the request.

In literature, NDN has been widely adopted in mobile ad-hoc networks. For example, V-NDN [3] illustrates NDN's promising potential in providing a unifying architecture that enables networking among all computing devices independent from whether they are connected through wired infrastructure, ad-hoc, or intermittent DTN. A multihop and multipath VANET routing algorithm [4] is proposed for NDN, that exploits using multiple paths concurrently to enable faster content retrievals. To better support the cases that vehicle location changes fast, CCLF [5] is a novel forwarding strategy using both location information and content connectivity to decide NDN request forwarding. The existing approaches can well support the mobility and dynamicity of ad-hoc networks. However few works consider the need of an "in-line" routing and payment protocol to incentivise packet routing/forwarding. In an ad-hoc environment, the trust amongst nodes are weak and subtle. The inline approach enables token transfers in pace with packet routing. The main advantage is that the service consumer can timely detect misbehaving routes thus avoiding them, and in consequence minimising the monetary loss.

R2P2, using blockchain as the settlement platform, realises the "in-line" payment function. Blockchain enables tokenising packet routing services. Different from centralised $3^{rd}$-party organisations, blockchain as a settlement platform, at the same time possesses the multiple advantages of neutrality, barrier-free accessibility and robustness. Using a public blockchain such as Ethereum, any nodes can register an account and transfer tokens anytime ad anywhere without worrying about single node failures and tampered ledgers. However, blockchain due to si architecture limitation, has serious performance issues to support concurrent massive transaction. To this end, this paper considers using off-chain payment channels to handle massive and high-frequency transactions thus reducing interaction with blockchain. The key contribution are the following:

- *A price-aware routing protocol*: this protocol allows consumers to explore the available routes and estimate the cost of using each.

- *A hybrid forwarding strategy*: to adapt to time-varying connectivities, routers switch amongst 3 different forwarding strategies to enhance network performance.
- *A proof-of-forwarding scheme*: the actual forwarding plane is fully controlled by routers, while malicious routers may direct packets to undesirable paths. The proof reveals the authenticity of a forwarding path.
- *An off-chain payment method*: it employs off-chain payment channels to enable charging routing services and blockchain as settlement platform.

## II. NAMED DATA NETWORK

NDN defines 2 types of packets [6]: Interest and Data. These packets are named by a Uniform Resource Identifier (URI). A content consumer sends Interest packets to fetch content. Once an Interest reaches a data producer, the producer put the corresponding content segments into a Data packet and returns it to the consumer. In this way, the content is not tied to any host. This address-less routing design enables content transmission between a consumer and any node that owns the data via any paths. This features is very useful to mobile ad-hoc network. For example, content can be distributed (cached) in the ad-hoc network so even if the original producer loses its connection, the consumer can still retrieve the content. In this paper, we refer *upstream* to the direction of Interset forwarding towards content and *downstream* to the direction of Data forwarding towards consumers.

## III. PRICE-AWARE ROUTING PROTOCOL

The proposed price-aware routing protocol features content consumers the ability to acquire the routes to accesing content and the cost of using each route. In essence, the protocol is developed to answer two fundamental questions: 1) how to route and 2) how much to pay. The two answers will be acquired via route discovery. Similar to the conventional ad-hoc routing protocols for NDN, the proposed protocol employs broadcasting to discover routes and collect the price of using each. The steps are as follows, with an example in Figure 1.

Each Interest packet contais a new field called *hop_info* which contains two identifiers: *local* and *remote*. Any Interest packet with *remote* set to *NULL* will be broadcasted until it reaches the content.

1. Initially, the consumer (Node A) sets *local* in the Interest packet to its local MAC address (00-14-XX) and broadcasts it. Since the consumer does not know any route to access content, *remote* is set to *NULL*.
2. Any router (Node B) received a broadcasting Interest packet will create a PIT entry to record the previous hop (Node A) with the address (00-14-XX). Then, the router changes *local* to its own MAC address (00-40-XX), which helps neighbours creat PIT entries.
3. After the Interst packet is received at the content provider (Node C), a corresponding route discovery Data packet is generated. The Data packet carries two fields in which one is *route* initialised as an empty stack (FILO), and the other one is *price*. The provider sequentially pushes its own address and the downstream (Node B's) address (according to the PIT entry) into *route*, i.e. the $1^{st}$ element is the downstream address (00-40-XX) and the $2^{nd}$ element is the local address (00-30-XX). Then, it puts the cost (15u) into *price*. Then the Data packet is sent to the network.
4. Once a router (Node B) receives a Data packet from an upstream node, the router creates an FIB entry for this Data or for the corresponding prefix. The FIB entry sets the upstream address (00-30-XX, Node C) as the next hop. Meawhile, FIB maintains a sliding-window of *price* for the next hop. It allows the router to estimate the minimum cost of using this next hop. R2P2 needs this value for forwarding (Section IV).
5. Then, the router (Node B) checks whether the $1^{st}$ element of *route* (00-40-XX) matches its own address. If not, the Data packet is discarded i.e. the Data should be returned by another router. Otherwise, the router looks into PIT and find the downstream to return it. In order to do so, the router pushes the downstream address (00-14-XX) into *route*. Also, the node increase *price* by adding its cost (12u + 3u = 15u).
6. The Data packet will finally arrives at the consumer (Node A), in which *route* reveals a route to access the content and *price* shows the cost of using the route.

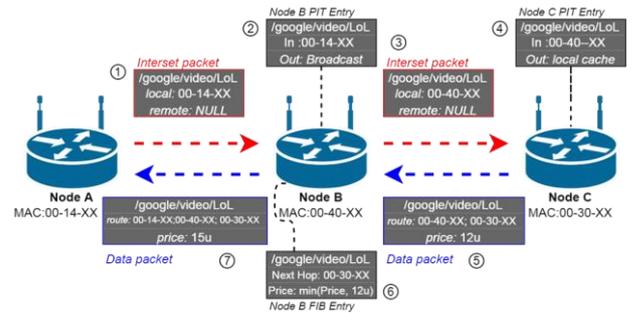

Figure 1 Broadcasting Example for Route Discovery

Using broadcast to discover routing paths brings the risk of creating broadcast storm. Extensive research works have been proposed to mitigate this issue in NDN. R2P2 supports any existing approach e.g. [7] to realise the broadcasting strategy.

In order to react to the changing topology timely, each router periodically tests the connectivity from itself to its neighbours. In particular, the router maintains a *keep alive* timer for each neighbour. Every given time period (e.g. 0.1s), each router broadcasts a *keep alive* signal to its neighbours to reset their timers. If the timer of a neighbour is not reset in time so timed out, the router considers this neighbour to be disconnected i.e. it disables all the relevant FIB entries that regards this neighbourhood as the next hop, until a new keep alive signal is received.

## IV. HYBRID FORWARDING STRATEGY

Based on the routes obtained through broadcasting, routers distribute Interst packets to retrieve content via adaptive forwarding. These Interest packets are requested by consumers who put the corresponding payment inside each packet. The detailed design will be presented in Section VI. For Interest packet forwarding, R2P2 proposes a hybrid approach which allows a router to switch amongst 1) source routing, 2) broadcasting and 3) minimum cost, adapting to the changing network statuses. The need of such a strategy originates from three goals: 1) improving consumer utility; 2) supporting new connectivities; 3) adapting to intermeditent connections. The strategy can be summarised as:

*"Routers selectively broadcast Interest requests to discover emergin routes without jamming the network. Consumers decide the route per request to control their utility. In the case that the route specified by a consumer fails, routers switch to back-up (minimum cost) routes to re-route the request."*

*1) Source routing:* it allows consumers to direct Interest packets to their preferred routes to download content. It requires each Interest packet to carry *route* so that routers know how to forward it. Source routing can bring a significant benefit to consumers because the consumer can measure the service quality of a known path and calculate its monetary cost (*price*). In consequence, the consumer can optimise its local utility by weighing cost and service quality.

*2) Minimum Cost*: Due to unstable connectivitys, some routes may become unavailable occasionally or permanently. If routers discard all the Interest packets that fail in source routing (e.g. the specified next hop is not alive), the service quality will be affected. As a back up, routers will forward these Interest packets to the hop with the minimum cost. Because giving a fixed amount of tokens paid by a consumer, the route with the minimum cost is most likely to serve the content, i.e. higher costs may cause packets to be rejected by intermediate routers.

*3) Broadcasting:* In the worst case, a router may find that all its next hops become failed for the incoming Interest packets. In this case, the router can selectively broadcast a few of the packets to re-discover routes meanwhile the other packets are discarded or NACKed. A round-robin method can be used to support the case that multiple flows all lose their routes. Note that R2P2 never uses broadcasting to deliver content for the following reasons: 1) the repeat deliveries of the same content files causes significant bandwidth overheads and 2) the monetary costs for content delivery via broadcasting routes are unpredictable and uncontrollable to consumers.

## V. Proof-of-Forwarding

During source routing, *route* is an instruction to direct packet forwarding. Nevertheless, dishonest nodes are capable to forward Interest packets via their routes. To this end, the consumer needs to prove that an Interest packet is always forwarded via the consumer-specified path. R2P2 proposes a Proof-of-forwarding (PoF) scheme to check the integrity of the forwarding path. In general, PoF requires routers to iteratively attach digital signatures to Data. If certain routers are not involved in forwarding, the consumer can detect it as some signatures will be missing. If unrelated nodes participate packet forwarding and do not attach digital signature, they will be considered lower layer devices and will not receive payments. If they make any change to the content or to the other digital signatures, the change can be easily detected by the consumer.

*A. Scalability Issue*

Similar to path and price probing, PoF requires routers to attach digital signatures to Data packets. If the PoF is performed at the packet level, i.e. each router signs every forwarded Data packet, PoF may become the throughput bottleneck of the whole forwarding process. According to a simple experimental test, a moderate CPU (e.g. Intel i5-

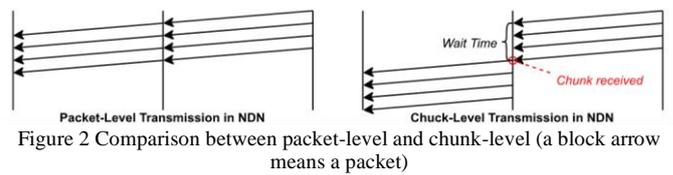
Figure 2 Comparison between packet-level and chunk-level (a block arrow means a packet)

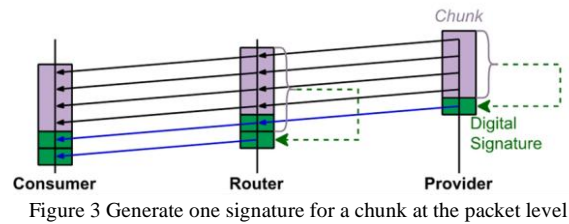
Figure 3 Generate one signature for a chunk at the packet level

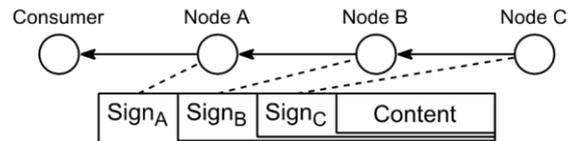
Figure 4 data structure of a signed content chunk

2450M at 2.50GHz, using Python3.7.4 ECDSA library) can sign and verify at most 1250 packets (1500Byte) per second. In this case, PoF can only support at most 15Mbps bandwidth, means the CPU is fully working on PoF.

*B. Scalability Solution*

We consider two methods to improve throughputs. The 1st one is to deploy NDN as an overlay. For example, if NDN is running on top of UDP, each Data packet can contain up to 65,507Byte (considers 28Byte UDP/IP headers). Considering the additional computing latency to signer a larger packet, the maximal bandwidth is around 550Mbps, using the same CPU. Although 550Mbps is sufficient for end-user applications (e.g. 4K video streaming), the high computing load of the CPU will quickly drain the battery.

It may be argued that NDN can run at chunk-level (the size of a chunk can be at a megabyte level), which easily extends the throughput to be higher than 10Gbps. However, chunk-level transmission introduces a high transmission delay. Due to the one-Interest-one-Data fashion in NDN, the chunk level transmission means that each NDN node needs to first fetch the whole content chunk then forward it to the downstream nodes (towards the consumer), as shown in Figure 2. We can see that the chunk-level transmission significantly increases latencies due to the long wait of receiving the full chunk at each hop.

Instead of running NDN at the chunk level, our proposed 2nd method is to generate a PoF signature for every *N* packets. In this case, the consumer needs to specify the Data packets belonging to the same group to travel through the same path. For example, the consumer can specify a video chunk (2MB) to be delivered via a single path, as shown in Figure 3. The Interest and Data packets are still forwarded at the packet-level. The signature is generated once a node finishes downloading the whole chunk. The delay is minimised because once a node receives a Data packet, it will forward the packet to downstream. Meanwhile, the node caches the Data packets belonging to the same chunk until all packets are received (and signed) or the pending Interests are all timeout. The data structure of a signed chunk is given in Figure 4.

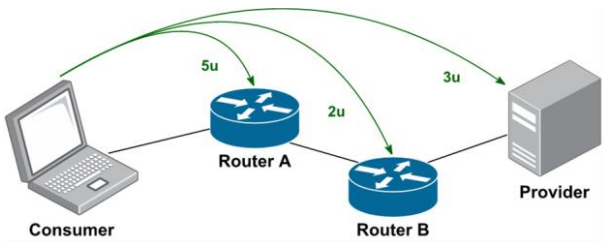

Figure 5 Consumer pays every node

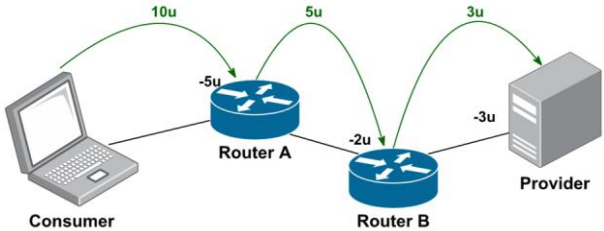

Figure 6 Hop-by-hop payment

Once a router receives a content chunk (with the signatures from previous nodes), using its private key the node generates a new signature for both the content chunk with the previous signatures. The consumer can verify the forwarding path by iteratively checking signatures using the public keys. Because the signatures are generated in a chained manner, a malicious node cannot fabricate a valid path thus to gain the trust. However, a malicious node is capable to tamper content and/or signatures thus causing it to be invalid. This malicious behaviour can be easily identified by the consumer and the neighbour nodes. As a result, the consumer will discard the tampered chunk and proactively terminate the contract and switch to other trustable paths. Even though this design requires the consumer to download all packet of a chunk from the same path, the multipath transmission can still be enabled at a different level, i.e. the consumer can concurrently download and pay for other chunks via alternative paths/sources.

## VI. OFF-CHAIN PAYMENT

The proposed in-line payment requires consumers to pay routers and providers frequently thereby mitigating debt repudiation. Specifically, R2P2 requires consumers to pay each delivery by adding tokens in Interest packets. So that if any router does not receive sufficient tokens, it will reject forwarding, and if a consumer finds a path not working, it will no longer pay for it.

In practice, the transaction throughput of the existing blockchain platform is quite limited (e.g. Ripple XRP can only support around 1500Tps), therefore it is impractical to support frequent token transfer between routers via on-chain transactions. In contrast, off-chain payment is a more scalable solution. This section compares two methods as given below:

*1) Consumer-pay-all:* The first method requires a content consumer to pay each router via a separated payment channel. An example is given in Figure 5. This approach has two main advantages. First, the consumer can accurately control the payment to each node. Second, routers do not need to deposit tokens to the payment channel. Nevertheless, this method is not scalable because every node may need to manage a number of channels in which each one corresponds to a consumer or a path, means to non-trivial transaction overheads to settle all the channels.

*2) Hop-by-hop payment:* The 2$^{nd}$ method lets a consumer pay routers hop-by-hop along each route. The connectivity amongst NDN nodes naturally forms a payment network therefore the consumer does not need to pay each node via a separated channel. Instead, the consumer can request the nearby node to pay remote nodes. An example is shown in Figure 6. A consumer would like to use the 3-hop path to download the content. Via *route*, the consumer can estimate the cost of using each hop (e.g. 5u, 2u, 3u). This allows the consumer to know the total cost (e.g. 10u) of using the whole path. For each payer (Consumer), it appends an off-chain payment (10u) into an Interest packet and sends it to the next-hop payee (Router A). The honest payee first deducts its income (e.g. 5u) from the payment and then forwards the "remaining" part with the Interest packet to the next payee (Router B). Finally, the Interest packet and the last payment will be received by the content provider.

In practice, this hop-by-hop payment needs the support of micropayment channels in which each hop corresponds to a channel. In general, a micropayment channel enables the payer and the payee to set up a local funding pool. They can deposit tokens and update the latest balances using a 2/2 multi-signature scheme, without calling blockchain. Finally, they can decide a time point to settle the balances using blockchain.

## VII. CONCLUSION

This paper demonstrated R2P2, aiming to incentivise network resource sharing among mobile nodes thereby acheiving content routing services. Hop-by-hop micropayment channels enable NDN nodes to transfer tokens thus allows consumers to pay for packet forwarding. The trust of forwarding packet along the specified path is realised via a proof-of-forwarding scheme. A chunk-level proof is proposed to guarantee scalability. Based on cost and measured performance, a consumer can calculate the utility value of each path and adjust the traffic on different paths thus to optimise its overall return. R2P2 is a new-born project. In the future, we are going to develop and implement a dynamic pricing scheme with R2P2 and evaluating the performance using real-world experiments.